\documentclass[aps,prl,twocolumn,groupedaddress]{revtex4}
\usepackage{graphicx}

\begin{document}
\newcommand{\beq}{\begin{equation}}
\newcommand{\eeq}{\end{equation}}
\newcommand{\bea}{\begin{eqnarray}}
\newcommand{\eea}{\end{eqnarray}}
\def\lsim{\:\raisebox{-0.5ex}{$\stackrel{\textstyle<}{\sim}$}\:}
\def\gsim{\:\raisebox{-0.5ex}{$\stackrel{\textstyle>}{\sim}$}\:}

\title{Planar Voronoi cells and the failure of Aboav's law}



\author{H.J. Hilhorst}
\email{Henk.Hilhorst@th.u-psud.fr}

\affiliation{Laboratoire de Physique Th\'eorique, B\^atiment 210,
Universit\'e de Paris-Sud, 91405 Orsay, France }

\date{\today}

\begin{abstract}
Aboav's law is a quantitative expression of the empirical fact that in planar
cellular structures many-sided cells tend
to have few-sided neighbors. 
This law is nonetheless violated in the most widely used
model system, {\it viz.} the Poisson-Voronoi tessellation. 
We obtain the correct law for this model: 
Given an $n$-sided cell, any of its
neighbors has on average $m_n$ sides where 
$m_n=4+3(\pi/n)^{-\frac{1}{2}}+\ldots$ in the limit of large $n$.
This expression is quite accurate
also for nonasymptotic $n$
and we discuss its implications for the analysis of experimental data. 
\end{abstract}

\pacs{}

\maketitle




Two empirical rules play a key role in studies of planar cellular systems:
Lewis' law and Aboav's law.
Both are statements about the statistics of a cell's
most prominent properties, {\it viz.} its area and its number of sides.
Lewis' law \cite{Lewis} 
says that the average area $A_n$ of an $n$-sided cell 
increases with $n$ as $A_n=a_0(n-n_0)/\lambda$,
where $a_0$ and $n_0$ are constants, and
$\lambda$ is the two-dimensional cell density. 
Aboav \cite{Aboav70} noticed that many-sided cells tend to have few-sided
neighbors and {\it vice versa.} He expressed this correlation in terms of 
the average $m_n$ of the
number of sides of a cell that neighbors an
$n$-sided cell. Aboav's law, 
also called the Aboav-Weaire \cite{Weaire74} law,
asserts that
\beq
m_n=a+\frac{b}{n}\,,
\label{eqnmn}
\eeq
where $a$ and $b$ are constants. This law is in widespread use
\cite{MAI93,JeuneBarabe98,MejiaRosalesetal00,CRR96,GGS87,Eliasetal97,
Lemaitreetal93,SMS94,EarnshawRobinson94,MTB02,ZLM99}  
in the analysis of experimental data on cellular structures.

In nature, planar cellular systems come in a wide
variety. They include biological tissues \cite{Lewis,MAI93,JeuneBarabe98},
polycrystals \cite{Aboav70},
cells formed by particles trapped at a water/air interface
\cite{MejiaRosalesetal00}, cells in surface-tension driven B\'enard convection 
\cite{CRR96}, in two-dimensional soap froths \cite{GGS87},
and in magnetic liquid froths \cite{Eliasetal97}.

Alternatively, the cellular structure may appear when the data are subjected
to the Voronoi construction \cite{Okabeetal00}.
Examples are hard disks on an air table \cite{Lemaitreetal93},
a binary liquid during late stage coarsening \cite{SMS94},
two-dimensional colloidal aggregation \cite{EarnshawRobinson94},
nanostructured cellular layers \cite{MTB02},
and studies of two-dimensional melting \cite{ZLM99}.

Many of these cellular systems,
in spite of all their diversity,
closely obey Lewis' law
from $n \approx 5$ up; also in many, Aboav's law 
appears to hold with good accuracy in
the full experimentally accessible range, {\it i.e.} from
$n=3$ to $n$ typically between $9$ and $12$. 
Fitting (\ref{eqnmn}) to
experimental data leads to values for the coefficient $a$ 
ranging from $4.6$ to $5.3$
and for $b$ from $7.0$ to $8.5$.
This observed similarity of behavior,
sometimes called ``universality,''
is generally attributed to the strong geometrical constraints that accompany
a division of the plane into convex cells; 
by contrast, 
physical or biological mechanisms are believed 
only to lead to corrections. 
It is of obvious interest to subtract from the experimental data
any purely geometrical effect that one can isolate
in order to identify the mechanisms at work that cause the corrections.

Although Lewis' and Aboav's laws arose initially
merely as good descriptions of the
available data, several ``derivations'' have since conferred to them
the status of theoretical truths valid for all $n$ and embodying
a purely geometrical theory.
Most attempts \cite{RivierLissowski82}
to derive these laws 
apply maximum entropy principles to a hypothesized entropy functional
\cite{EdwardsPithia94}.
A critique of this usage of the maximum entropy method  
is due to Chiu \cite{Chiu95}, who has shown that 
no firm conclusions can be drawn from it.
In any case, none of these derivations connects either Lewis' or Aboav's
law by a first principle calculation
to a microscopic model of planar cellular structure.
\vspace{1mm}

{\it First-principle approach.\,\,--}
The simplest microscopic model of a planar 
cell model is the Poisson-Voronoi (PV) tessellation: 
It is obtained by constructing the Voronoi cells
\cite{Okabeetal00} of a configuration of
randomly and uniformly distributed {\it point centers} in the plane,
for convenience often called ``seeds'' (but without
the implication that they are material). 
The PV tessellation 
is therefore a natural candidate
for the purely geometrical theory,
and certainly the easiest one to handle.

The statistical properties  of the PV tessellation
have been studied analytically and by Monte Carlo methods
(see Ref.\,\cite{Okabeetal00} for a review).
Nonetheless, a first principle derivation of expressions for $A_n$ and $m_n$
has long seemed forbiddingly difficult.
Numerically it is known \cite{BootsMurdoch83,LeCaerHo90,KumarKurtz93,LMH93} 
that Aboav's law is {\it not\,} exactly valid for 
the Poisson-Voronoi tessellation:
Whereas Eq.\,(\ref{eqnmn}) predicts $nm_n$ to be linear in $n$,
this quantity shows in fact a very small but distinct downward curvature 
for the PV tessellation.
To accommodate this discrepancy modifications of Aboav's law 
have been suggested, in particular, by Boots and Murdoch \cite{BootsMurdoch83}
and by LeCa\"er and Ho \cite{LeCaerHo90}. The former authors 
write $m_n=A+Bn^{-1}+Cn^{-2}$, which was later found to be true exactly
for a class of graphs in field theory \cite{GKY92}.

A recent letter \cite{HJHletter05} has achieved
a step forward in the statistical mechanics of planar Voronoi tessellations. 
It opens up the possibility for an 
expansion in powers of $n^{-\frac{1}{2}}$
of all quantities of interest related to the $n$-sided Voronoi cell; 
an immediate result was
that Lewis' law holds with coefficient 
$a_0=\frac{1}{4}$ for asymptotically large $n$.
\vspace{1mm}

\begin{figure}
\vspace{5mm}
\scalebox{.36}
{\includegraphics{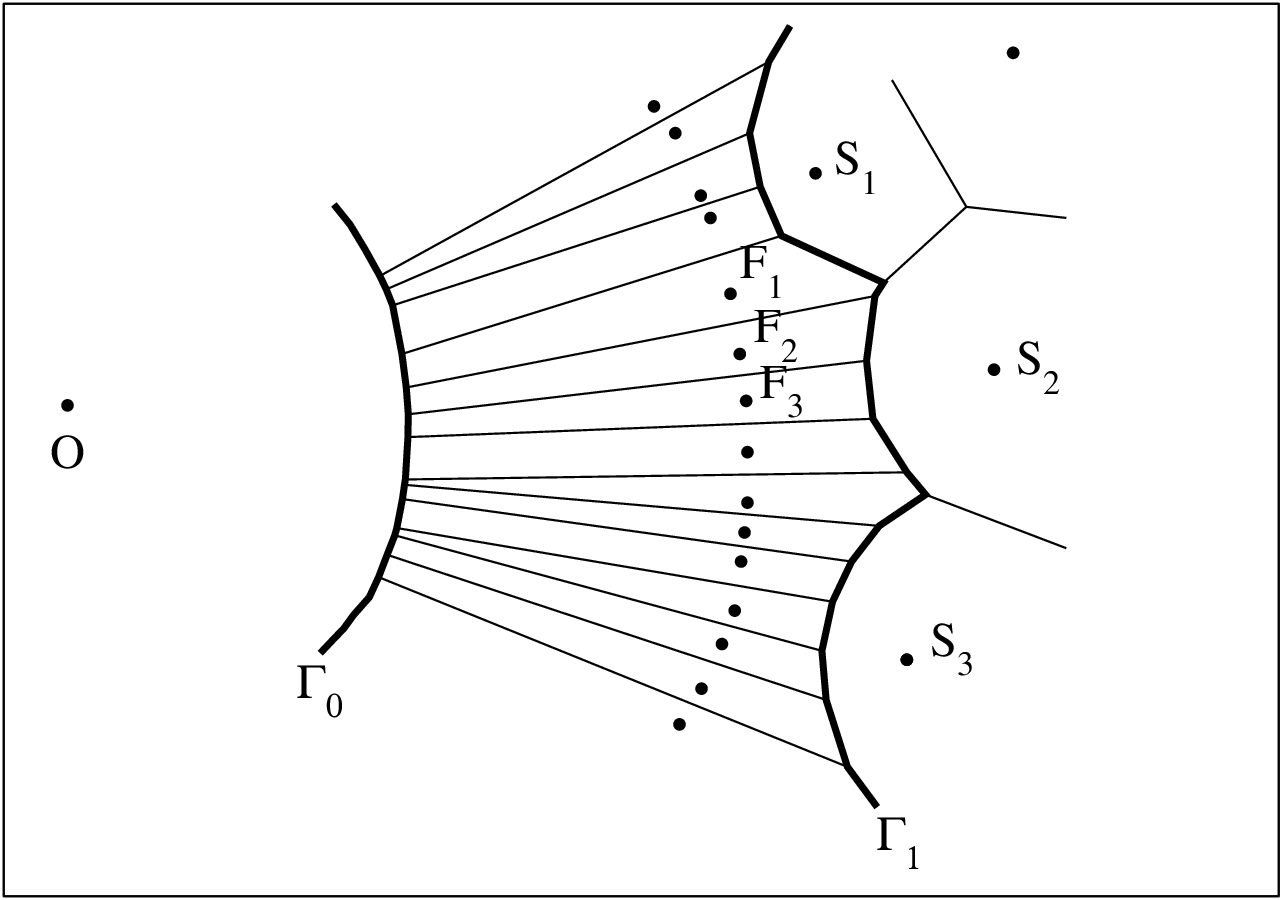}}
\caption{{\small 
Schematic picture of an $n$-sided Voronoi cell having $n \approx 100$
and containing a seed at $O$. First and second neighbor cells 
contain seeds at $F_1, F_2, F_3, \ldots$ and 
at $S_1,S_2,S_3,\ldots$, respectively.
All solid line segments separate Voronoi cells. Among these,
the heavy solid line $\Gamma_0$ is the perimeter of the $n$-sided cell,
which is close to circular.
The heavy solid line $\Gamma_1$  separates the first from the second
neighbors. Both are piecewise linear on a scale $n^{-\frac{1}{2}}$. On the
scale of 
order $1$ the incipient piecewise parabolic structure of $\Gamma_1$ 
is discernible.}} 
\label{fig1}
\end{figure}
{\it Aboav's law.\,\,--}
Whereas Lewis' law refers to a single cell, Aboav's deals with the
intrinsically more difficult problem of correlations between cells.
Here we track down the implications of the 
$n^{-\frac{1}{2}}$ expansion
for Aboav's law.
We consider an $n$-sided cell with $n$ very large.
Cells with very many sides are extremely
rare, but {\it if\,} one occurs, then its environment
must look as depicted in Fig.\,\ref{fig1},
where the 
$n$-sided cell of a ``central'' seed at $O$
is surrounded by $n$ 
strongly elongated first neighbor cells containing seeds $F_i$.
Independent evidence for such a geometry comes from work by 
Lauritsen {\it et al.} \cite{LMH93}, 
who Monte Carlo simulated a Hamiltonian favoring the
appearance of many-sided cells. 

In Fig.\,\ref{fig1} at least four different length scales play a role, each of
them proportional to its own power of $n$. 
Heavily relying on the work of Ref.\,\cite{HJHletter05}, which for the present
purpose we extend and interpret,  
we establish the following list of scales: 

(i) The perimeter $\Gamma_0$ of the central cell typically runs within 
an annulus of center $O$, of radius $R_{\rm c} =
(n/4\pi\lambda)^{\frac{1}{2}}$, and of width of 
order $1$. Hence for $n\to\infty$
the perimeter tends towards a circle of radius $R_{\rm c}$.
Consequently in that limit the first neighbors $F_i$ will be 
on a circle of radius $2R_{\rm c}$. 

(ii) Two successive vertices 
on the perimeter of the central cell have an 
average distance 
$\ell_{\rm vert}=2\pi R_{\rm c}/n=(\pi/n\lambda)^{\frac{1}{2}}$. Consequently,
two successive first neighbor seeds 
$F_i$ and $F_{i+1}$ have an average distance $2\ell_{\rm vert}$.

(iii) Locally the positions of the vertices of the central cell
are strongly aligned: Their radial coordinates
have rms deviations 
of order $n^{-\frac{3}{2}}$ with respect to the
locally averaged radius.   
For the purpose of the present discussion we 
may set these deviations equal to zero; their smallness implies that for
$n\to\infty$ the perimeter of the central cell 
becomes a smooth curve.
Similar statements hold for the curve, not drawn in the Fig.\,\ref{fig1},
that links the successive first neighbors.

(iv) In the region to the right of the heavy solid line 
$\Gamma_1$, which is occupied by 
second and further neighbors, the 
seed density has its ``background'' value $\lambda$,
{\it i.e.} is of order $n^0$ \cite{footnote1}. 
\vspace{1mm}

On the basis of this picture we reason as follows.
The perimeter of the central cell
carries a vertex line density
$\rho_{\rm vert}=1/\ell_{\rm vert}=(n\lambda/\pi)^{\frac{1}{2}}$, 
which for $n\to\infty$ tends to infinity. 
In spite of this diverging line density the surface 
density $\lambda$ of the seeds 
to the right of the curve $\Gamma_1$
stays of order $n^0$. It follows that the central cell will have
$\sim n^{\frac{1}{2}}$ 
second neighbor cells and that each second neighbor $S_j$ will 
be adjacent to $\sim n^{\frac{1}{2}}$ first neighbor cells $F_i$
(where we call the cells by the names of their seeds).
Fig.\,\ref{fig1} shows that under these circumstances each first
neighbor $F_i$ is most likely to have itself four neighbors, 
{\it viz.} the central $n$-sided cell, a single second neighbor cell, 
and two other first neighbors, $F_{i-1}$ and $F_{i+1}$.

We now focus on the exceptional $F_i$ that have five neighbors
due to their being adjacent to {\it two\,} second neighbors 
$S_j$ and $S_{j+1}$. 
An example is the cell marked $F_1$ 
in Fig.\,\ref{fig1}, which is adjacent to both $S_1$ and $S_2$.
Let us denote by $f_5$ the fraction of first neighbors that are $5$-sided.
In view of the scaling relations that precede we expect that 
$f_5=c n^{-{\frac{1}{2}}}+\ldots$\,, where $\,c$ is a numerical
coefficient and the dots indicate terms of higher order in $n^{-\frac{1}{2}}$. 
Any $6$- and higher-sided $F_i$ will contribute only to these dot terms. 
Hence we have
\bea
m_n &=& 4\,(1-f_5) + 5f_5\nonumber\\[2mm]
&=& 4 + c n^{-{\frac{1}{2}}} + \ldots\,.
\label{relmnc}
\eea
\begin{figure}
\vspace{5mm}
\scalebox{.34}
{\includegraphics{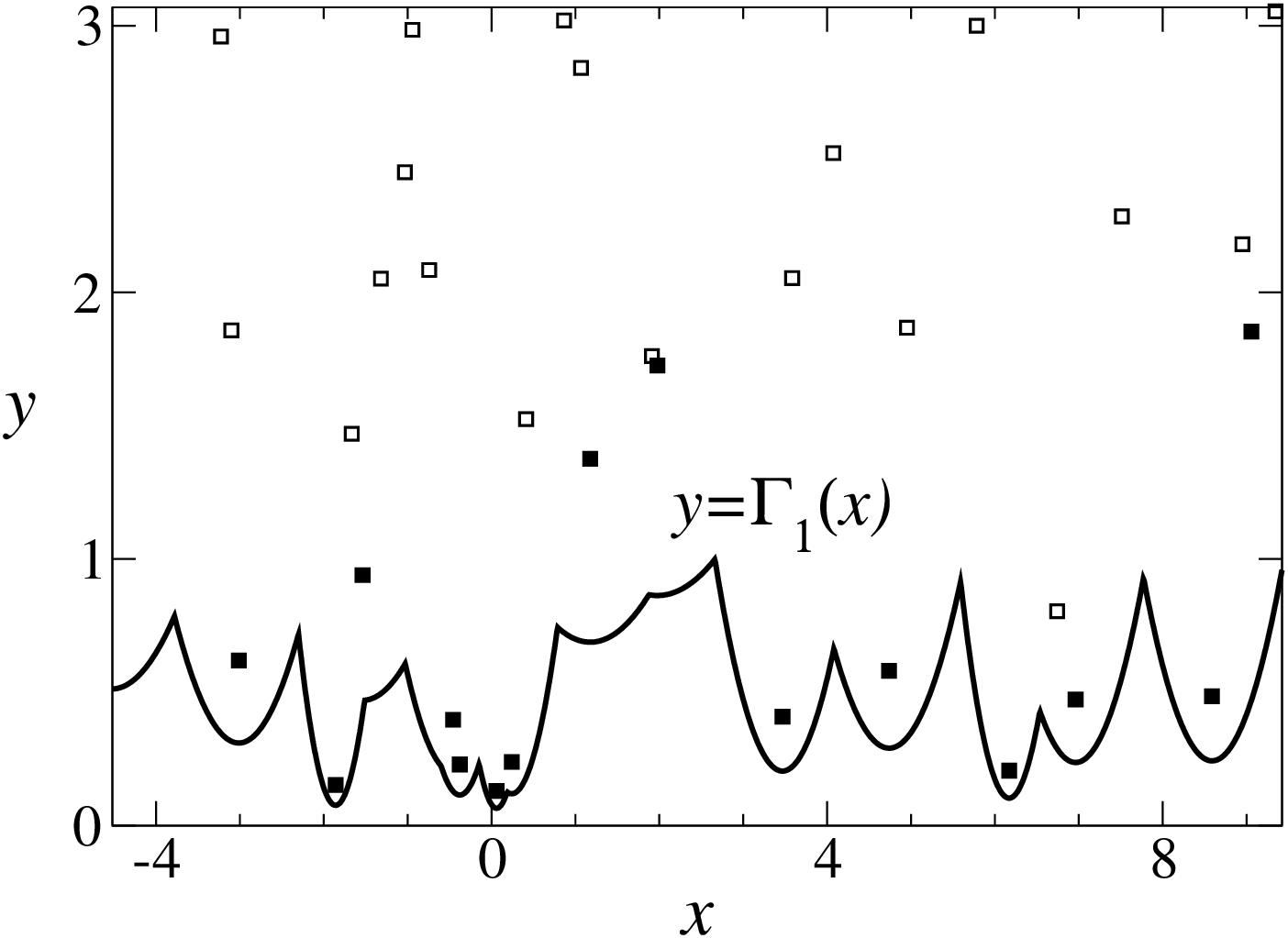}}
\caption{{\small The $x$-axis represents a continuum of first neighbors.
    The upper half-plane is randomly filled with seeds of
    uniform density $\lambda$ (here $\lambda=1$). 
    The piecewise parabolic curve $y=\Gamma_1(x)$
    separates the region of the half-plane
    closer to the $x$ axis than to any of the random seeds from its
    complement. The abscissae of the cusps of $\Gamma_1$ have a density of
    $\frac{3}{2}\lambda^{\frac{1}{2}}$ on the $x$ axis. 
    Solid and open squares represent second and further neighbors,
    respectively. Note that the horizontal and vertical scales are different.}}
\label{fig2}
\end{figure}
We next consider the solid line $\Gamma_1$ in Fig.\,\ref{fig1}, 
which separates the central seed's first from its second neighbors.
In the large $n$ limit the curve linking successive first neighbor seeds 
becomes a circle which may locally be replaced with a straight line.
In Fig.\,\ref{fig2} this same straight line is represented by the $x$ axis 
and the region of space containing the
second and further neighbors by the half-plane $y>0$.
The second and further neighbor seeds are uniformly distributed in the
upper half plane with the background density $\lambda$.
Since the first neighbors $F_i$ are dense on the $x$ axis,
the curve $\Gamma_1$
now divides the half-plane $y>0$ into a lower part 
of points closer to the $x$ axis than to any of the 
seeds, and its complement. 
Hence the function $y=\Gamma_1(x)$ is piecewise parabolic;
its incipient parabolic segments are discernable in Fig.\,\ref{fig1}.
To each cusp of $\Gamma_1(x)$ corresponds a $5$-sided first neighbor cell. 
Let $\ell_{\rm cusp}$ be the average distance between the abscissae
of two successive cusps of $\Gamma_1(x)$. 
Then it is clear that $f_5=2\ell_{\rm vert}/\ell_{\rm cusp}$. 
The determination of $\ell_{\rm cusp}$ for given seed density $\lambda$ in
the upper half plane is a problem in statistics that 
yields $\ell_{\rm cusp}=2/(3\lambda^{\frac{1}{2}})$
\cite{HJHunpublished}.
Using the expression for $\ell_{\rm vert}$ found above we
therefore have that $f_5=2(\pi/n\lambda)^{\frac{1}{2}} \times
3\lambda^{\frac{1}{2}}/2=3(\pi/n)^{\frac{1}{2}}$, 
whence $c=3\pi^{\frac{1}{2}}$.
Substituting this in (\ref{relmnc}) we conclude that for
the Poisson-Voronoi tessellation $m_n$ is exactly given by
\beq
m_n=4+3\sqrt{\frac{\pi}{n}}+\ldots, \qquad  n\to\infty.
\label{mnaspt}
\eeq
The inverse square root decay 
and the limiting value $m_\infty=4$ of Eq.\,(\ref{mnaspt})
are in contradistinction to Aboav's law (\ref{eqnmn}).\, 
This equation explains for the first time 
the downward curvature observed 
\cite{BootsMurdoch83,LeCaerHo90,KumarKurtz93,LMH93} 
in the $nm_n$ {\it vs.} $n$ curves for the PV tessellation.
\vspace{1mm}

\begin{figure}
\vspace{5mm}
\scalebox{.33}
{\includegraphics{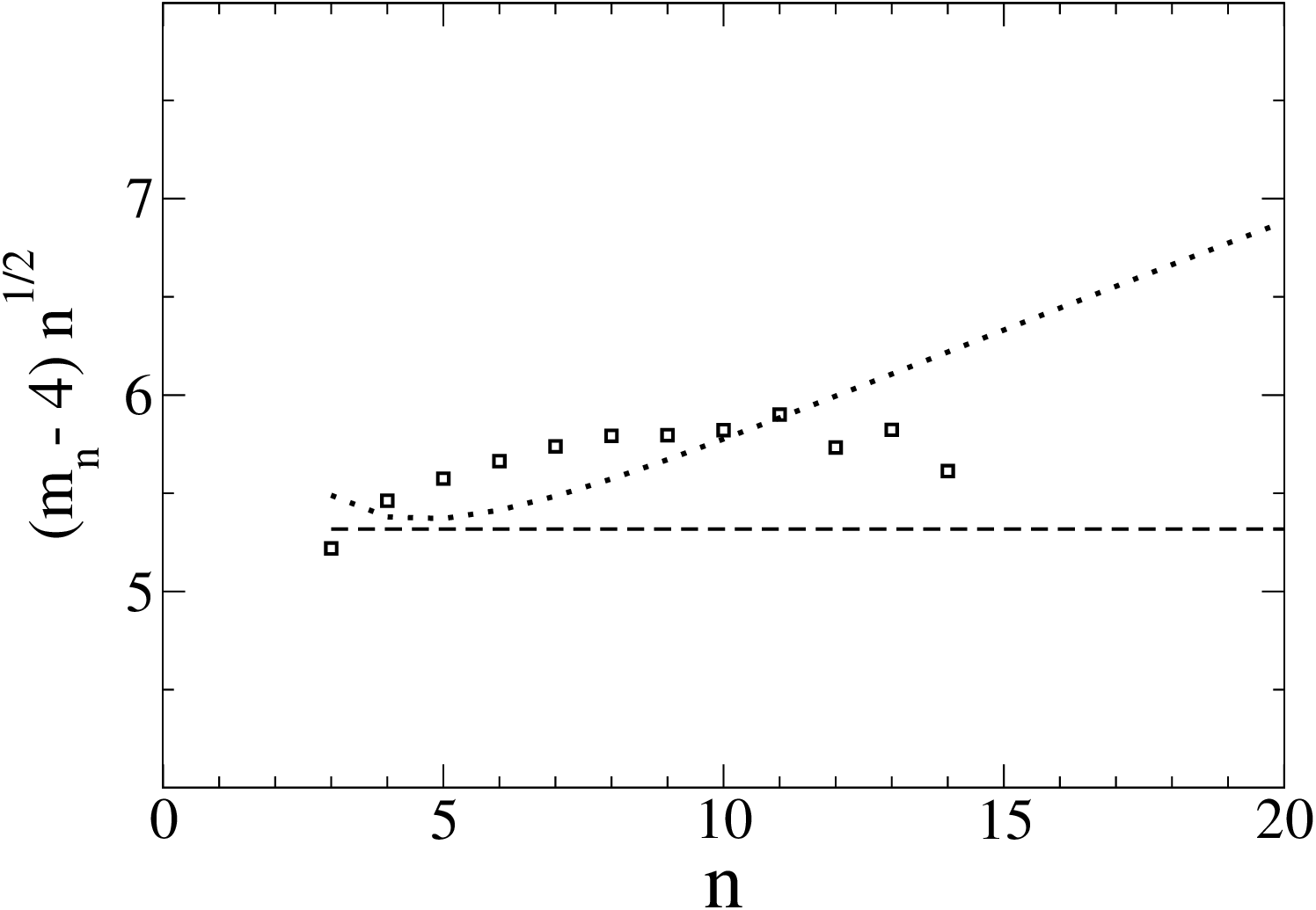}}
\caption{{\small 
     The two-cell correlation $m_n$ as a function of $n$.
     Open squares: Simulation values due to Boots and Murdoch
     \cite{BootsMurdoch83}.
     Dotted line: Best two-parameter fit with Aboav's law (\ref{eqnmn}).
     Dashed line: Our zero-parameter exact asymptotic result (\ref{mnaspt}),
     which in this representation is a constant equal to
     $3\pi^{\frac{1}{2}}$.}}  
\label{fig3}
\end{figure}

{\it Data analysis.\,\,--}  We cannot assess {\it a priori}
the applicability of Eq.\,(\ref{mnaspt}) to
$m_n$ data in the range of $n$ covered by
experiments and simulations. 
We have plotted in a new way in Fig.\,\ref{fig3} the
Monte Carlo data for $m_n$ due to Boots and Murdoch \cite{BootsMurdoch83}. 
The dotted curve shows
the best two-parameter fit to the data provided by Aboav's law 
(\ref{eqnmn}); it is
obtained for $a=5.251$ and $b=5.755$ \cite{Okabeetal00} and is generally
taken as evidence that this law fails for the PV tessellation.
The dashed line in
Fig.\,\ref{fig3} represents the right hand side of Eq.\,(\ref{mnaspt})
with the dot terms neglected. 
It appears that in the regime of the data our 
zero-parameter asymptotic result is numerically only slightly more off 
than Aboav's two-parameter law \cite{footnote2}.
This high degree of accuracy is remarkable in view of the fact that 
Eq.\,(\ref{mnaspt}) rests on an expansion around the
improbable event of a large $n$-sided cell.
But more importantly, there is full compatibility between our dashed line
and the downward trend of the data for $n$ larger than about 10.
One may attempt to improve our Eq.\,(\ref{mnaspt}) by including
higher order terms in $n^{-\frac{1}{2}}$ in the series, 
the simplest possibility being
\beq
m_n=4+3\sqrt{\frac{\pi}{n}} + \frac{d}{n},
\label{oneparam}
\eeq
where $d$ is adjustable.
However, there is no reason to expect that a
truncated expression, whether (\ref{oneparam}), (\ref{eqnmn}), or yet
another one, will yield the exact $m_n$ for all $n$.
\vspace{1mm}

The PV tessellation serves as 
a model of reference for general cellular systems
much in the same way as an ideal gas does for real gases.
The inverse square root decay of its two-cell correlation
rules out Aboav's law (\ref{eqnmn}) and
may be viewed as defining a universality class.
We cannot exclude that
someday some other microscopic model will be shown to
asymptotically obey Aboav's law (\ref{eqnmn}) to leading order in $n^{-1}$.
Such a model would then define a second universality class.
It is unknown at present which experimental systems are in the PV 
universality class.

Many different factors may potentially
lead to departures from PV-like behavior,
if not to a change of universality class. {\it E.g.,}
(i) in some cellular structures 
the ``seeds'' represent actual particles or larger
physical entities with mutual interactions;
(ii) a general planar cell structure cannot be derived from a set
of point centers by means of the Voronoi construction; and
(iii) some systems, like soap froths, are not in equilibrium but rather
in a -- supposedly scale invariant -- coarsening state. 
One may now hope for such cases to be able to discuss the deviations from 
PV tessellation statistics perturbatively.

To show how such a discussion might proceed,
we consider briefly 
and heuristically
an example from class (ii) above, 
{\it viz.} the Voronoi tessellations 
associated with hard core particles of finite diameter $a$.
The preceding analysis of the large $n$ limit
goes through as long as $n \lesssim n^*$, where
the crossover value $n^*$ is 
determined by the condition that the distance between adjacent 
first neighbors become comparable to the particle diameter. This gives
$2\pi R_{\rm c} \sim n^*a$, whence $n^* \sim \pi/(\lambda a^2)$.
For $n \gsim n^*$ the repulsion between the particles combined with the
condition that they be locally aligned 
imposes that the radius of the central cell grows as $R_{\rm c} \sim na$
and hence that $f_5$ must saturate at a value
$f_5 = c_0 a\lambda^{\frac{1}{2}}$ where $c_0$ is an unknown numerical
coefficient. 
Upon assuming that this knowledge may be expressed
with the aid of a scaling function 
${\cal M}$ we obtain for hard core particles of diameter $a$ 
the relation
$m_n = 4 + a\lambda^{\frac{1}{2}}{\cal M}(n\lambda a^2)$, 
valid in the scaling limit 
$n\to\infty$, $\lambda a^2\to 0$ with $x=n\lambda a^2$ fixed,  
and where the scaling function ${\cal M}$ satisfies ${\cal M}(0)=c_0$ and
${\cal M}(x) \simeq 3(\pi/x)^{\frac{1}{2}}$ for $x\to\infty$.
Hence for hard core particles the limiting value $m_\infty$
of (\ref{mnaspt}) is changed, {\it but the $n^{-\frac{1}{2}}$ decay law
remains}.  
Future work will have to deal with
this and other instances of deviations from PV statistics.
\vspace{1mm}

{\it Conclusion.\,\,--}
The simplest and most widely used model of a cellular
structure -- the Poisson-Voronoi tessellation -- 
does not obey the most widely used law -- Aboav's law (\ref{eqnmn}) --
governing the two-cell correlations. 
We have shown that for the PV tessellation
the $n^{-1}$ decay of Aboav's law should in fact be
replaced with an $n^{-\frac{1}{2}}$ decay. 
Experimental data of sufficient precision and/or covering a large enough 
range should be able to distinguish between the two and thereby
shed light on the
question of the universality classes and of the underlying cell
formation mechanisms.
We therefore advocate that experimental results be analyzed not only by 
Aboav's two-parameter fit (\ref{eqnmn}), but also by the simpler one-parameter
formula (\ref{oneparam}), of which the first two terms have a firm
theoretical basis.
\vspace{1mm}

{\it Acknowledgment.\,\,--}
This work was made possible by a six month sabbatical period (CRCT)
granted to the author by the French Ministry of Education.



\begin{thebibliography}{10}
\bibitem{footnote1} Fig.\,\ref{fig1} is schematic in that a faithful
  representation of all proportions would lead to an about twice
  higher density of the
  second and further neighbors to the right of $\Gamma_1$; as it is,
the picture better brings out the emergence
of the piecewise parabolic segments of $\Gamma_1$.
\bibitem{footnote2}
A best fit of Eq.\,(\ref{relmnc}) without its dot terms
to the Monte Carlo data yields a coefficient 
$c \approx 5.7$, only slightly higher than the exact value 
$c = 3\pi^{\frac{1}{2}}=5.317...$\,.
\end{thebibliography}
\end{document}